\documentclass[11pt,letterpaper]{article}
\usepackage[margin=1in]{geometry}

\usepackage{amsmath,amssymb,amsthm}
\usepackage{xcolor}
\usepackage{cite}
\usepackage{tikz}
\usetikzlibrary{arrows.meta,positioning}
\usepackage{tabularx}
\usepackage
[
colorlinks=true,
citecolor=blue,
linkcolor=blue,
urlcolor=blue
]{hyperref}
\usepackage{microtype}
\usepackage{booktabs}
\usepackage{array}
\usepackage{stmaryrd}

\newtheorem{theorem}{Theorem}
\newtheorem{lemma}[theorem]{Lemma}

\theoremstyle{definition}
\newtheorem{definition}[theorem]{Definition}
\newtheorem{remark}{Remark}[section]

\newtheorem{construction}{Construction}[section]

\newcommand{\F}{\mathbb F}
\newcommand{\cA}{\mathcal A}
\newcommand{\cB}{\mathcal B}
\newcommand{\cC}{\mathcal C}
\newcommand{\cG}{\mathcal G}
\newcommand{\cN}{\mathcal N}
\newcommand{\cP}{\mathcal P}

\newcommand{\distE}{d_{\mathrm E}}

\newcommand{\id}{\operatorname{id}}
\newcommand{\good}{\operatorname{Good}}
\newcommand{\rep}{\operatorname{rep}}
\newcommand{\ser}{\operatorname{ser}}
\newcommand{\ceil}[1]{\left\lceil #1\right\rceil}
\newcommand{\floor}[1]{\left\lfloor #1\right\rfloor}

\newcommand{\abs}[1]{\left|#1\right|}
\newcommand{\nz}[1]{\llbracket#1\rrbracket}

\title{Explicit Edit Codes via Multicoloring}
\author{Huiling Xie\thanks{School of Science and Engineering, The Chinese University of Hong Kong, Shenzhen, Guangdong 518172, China. Email: \texttt{huilingxie@link.cuhk.edu.cn}.}
	\and Zitan Chen\thanks{School of Science and Engineering and Future Networks of Intelligence Institute, The Chinese University of Hong Kong, Shenzhen, Guangdong 518172, China. Email: \texttt{chenztan@cuhk.edu.cn}.}}
\date{}
\hypersetup{pdftitle={Deterministic Edit Codes via Enumerative Multicoloring},pdfauthor={Huiling Xie and Zitan Chen}}
\begin{document}
	\hypersetup{pageanchor=false}
	\maketitle
	\begin{abstract}
		For any fixed integers $q\ge2$ and $t\ge1$, we construct explicit $q$-ary codes of length $n$ that correct any combination of at most $t$ insertions, deletions, and substitutions, with redundancy at most $2t\log_q n+O_{q,t}(\log\log n)$ $q$-ary symbols.
		This matches the leading redundancy term of Levenshtein's classical existential bound.
		Building on the standard graph-coloring approach, we use a prime-based multicoloring of the edit-confusion graph to assign multiple colors to each $q$-ary word of length $m<n$.
		Appending a short, edit-protected representation of an assigned color to the length-$m$ word yields a length-$n$ codeword.
		The choice among the assigned colors carries additional message information, thereby reducing the redundancy.
		The construction admits deterministic encoding and decoding	in $\widetilde O_{q,t}(n^{6t+2})$ and $\widetilde O_{q,t}(n^{7t+2})$ time, respectively.
	\end{abstract}
	\thispagestyle{empty}
	\newpage
	\setcounter{page}{1}\hypersetup{pageanchor=true}
	
	\section{Introduction}
	
	Fix integers \(q\ge2\) and \(t\ge1\), and let \(\cA=\{0,\ldots,q-1\}\).
	We consider a \(q\)-ary edit channel that applies any combination of at most \(t\) insertions, deletions, and substitutions to a transmitted word. Our goal is to construct codes \(\cC\subseteq\cA^n\) with low redundancy that admit deterministic polynomial-time encoding and decoding.
	Throughout, we measure redundancy in $q$-ary symbols of the code alphabet. In other words, a code \(\cC\subseteq\cA^n\) has redundancy \(n-\log_q|\cC|\).

	The classical greedy argument \cite{levenshtein1966} selects codewords whose pairwise edit distances exceed \(2t\), establishing the existence of codes with redundancy
	\(2t\log_q n+O(1)\).
	Its straightforward implementation, however, searches an exponentially large space of words and does not provide a polynomial-time encoder.
	Our main result achieves the same leading redundancy term with deterministic polynomial-time encoding and decoding.
	
	\begin{theorem}\label{thm:edit-main}
		Fix integers \(q\ge2\) and \(t\ge1\).
		There are deterministic encoding and decoding algorithms such that, for all sufficiently large \(n\), the encoder is an injective map
		\[
		C:\cA^k\longrightarrow\cA^n,
		\qquad
		k\ge n-2t\log_q n-O(\log\log n),
		\]
		whose image \(\cC:=C(\cA^k)\) corrects every combination of at most \(t\) insertions, deletions, and substitutions.
		Encoding and decoding take \(\widetilde O(n^{6t+2})\) and \(\widetilde O(n^{7t+2})\) time, respectively.
	\end{theorem}
	
	Our construction builds on the standard graph-coloring approach	to constructing large codes.
	We use a prime-based multicoloring of the confusion graph to assign multiple colors to each message prefix, with disjoint color sets for confusable prefixes.
	We then append a short, edit-protected representation of an assigned color to the prefix.
	The choice among the assigned colors carries additional message	information, thereby reducing the redundancy.
	
	\subsection{Related literature}
	
	Research on codes correcting insertions, deletions, and	substitutions dates back to Levenshtein
	\cite{levenshtein1966}.
	Building on the Varshamov-Tenengolts (VT) construction	\cite{varshamov1965codes}, he obtained binary codes correcting	a single edit with redundancy \(\log_2 n+O(1)\), which is asymptotically optimal.
	Subsequent work has sought low-redundancy constructions with efficient algorithms for multiple errors and larger alphabets.
	
	\paragraph{Deletion codes.}
	
	Deletion correction provides a useful starting point for edit correction.
	For fixed-length codes, correcting \(2t\) deletions also guarantees correction of any mixture of at most \(2t\) insertions and deletions.
	Since a substitution can be simulated by one deletion and one insertion, a \(2t\)-deletion-correcting code therefore corrects \(t\) edits.
	
	For fixed \(q\) and \(t\), Levenshtein's classical lower and upper bounds on the minimum redundancy of \(t\)-deletion-correcting codes have leading terms \(t\log_q n\) and \(2t\log_q n\), respectively
	\cite{levenshtein1966,levenshtein2002bounds}.
	Refined deletion-packing arguments have strengthened converse bounds in various parameter regimes; see, for example, \cite{kulkarni-kiyavash2013,fazeli-vardy-yaakobi2015,cullina-kiyavash2014}.
	
	On the constructive side, we first consider binary deletion codes. Guruswami and H{\aa}stad \cite{guruswami-hastad2021} gave explicit binary
	two-deletion-correcting codes with redundancy \(4\log_2 n+O(\log\log n)\), matching the leading term of the classical existential bound.
	For fixed \(t\ge2\), Brakensiek, Guruswami, and Zbarsky \cite{brakensiek2018} obtained redundancy \(O(t^2\log t\log n)\) with near-linear encoding and decoding.
	Cheng, Jin, Li, and Wu \cite{cheng-jin-li-wu2022} achieved \(O(t\log n)\) redundancy with deterministic polynomial-time algorithms.
	More precise leading coefficients of \(\log_2 n\) were obtained by Sima and Bruck \cite{sima-bruck2021}, Sima, Gabrys, and Bruck \cite{sima-systematic2020}, and Song, Polyanskii, Cai, and He \cite{song-polyanskii2022},	whose constructions achieve coefficients \(8t\), \(4t\),
	and \(4t-1\), respectively.
	
	Over nonbinary alphabets, Tenengolts \cite{tenengolts1984} gave a nearly optimal construction for a single deletion.
	For two deletions, Song and Cai \cite{song-cai2023} obtained redundancy \(5\log_q n+O(\log\log n)\) for fixed even \(q\), with near-linear encoding using a polynomial-size lookup table.
	Liu, Tjuawinata, and Xing \cite{liu2024} attained the same asymptotic redundancy for every fixed \(q>2\), with polynomial-time encoding.
	For fixed \(t\ge2\), Sima, Gabrys, and Bruck \cite{sima2020optimal2} constructed \(q\)-ary \(t\)-deletion-correcting codes with redundancy	at most
	\(4t(1+\epsilon)\log_q n+o(\log n)\).
	
	Complementing these constructive advances, existential bounds have also improved.
	Alon, Bourla, Graham, He, and Kravitz \cite{alon2024} improved Levenshtein's greedy lower bound on code size by a logarithmic factor.
	For every fixed \(q\ge2\) and \(t\ge2\), their result gives	deletion-correcting codes with redundancy $2t\log_q n-\log_q\log_2 n+O(1)$.
	Kong, Tamo, and Wei \cite{kong-tamo-wei2025} subsequently refined the multiplicative constant in the corresponding asymptotic size bound.
	While these improvements affect lower-order redundancy terms, the recent work \cite{en-gad2026} reduces the leading	coefficient, establishing the existence of \(q\)-ary \(t\)-deletion-correcting codes with redundancy at most
	\begin{equation}\label{eq:rub}
		(2t-1)\log_q n+O(\log\log n),
		\qquad t\ge2.
	\end{equation}
	
	\paragraph{Edit codes.}
	
	For correction of \(t\) edits, classical packing and greedy arguments give redundancy lower and upper bounds of \(t\log_q n-O(1)\) and \(2t\log_q n+O(1)\), respectively.
	Moreover, every converse bound for \(t\)-deletion correction also applies to \(t\)-edit correction, since deletions are	a special case of edits.
	
	For a single edit, the asymptotically optimal binary VT constructions have been complemented by	nearly optimal constructions over nonbinary alphabets \cite{cai2021,gabrys2023}.
	For two binary edits, Sun and Ge \cite{sun-ge2024} obtained redundancy \(6\log_2 n+O(\log\log n)\).
	Pi and Zhang \cite{pi-zhang2025} subsequently reduced the lower-order term to a constant, obtaining redundancy \(6\log_2 n+8\).
	
	For fixed \(t\ge2\), the construction of Sima, Gabrys, and Bruck \cite{sima-systematic2020}	also yields binary \(t\)-edit-correcting codes with redundancy \(4t\log_2 n+o(\log n)\).
	More closely related to our approach, Li, Gabrys, and Farnoud \cite{li-gabrys-farnoud2025} developed a general framework for constructing error-correcting codes via distributed
	graph coloring.
	For fixed \(q\) and \(t\), their framework gives explicit \(t\)-edit-correcting codes with redundancy \(4t\log_q n+O(\log\log n)\) and polynomial-time encoding
	and decoding.
	On the existential side, \cite{en-gad2026} showed that the improved bound \eqref{eq:rub} holds for \(t\)-edit-correcting codes as well.
	
	\paragraph{Concurrent work.}
	
	While this paper was in preparation, Bui, Wang, and Vu \cite{bui2026binary} gave an explicit construction of binary \(t\)-edit-correcting codes with redundancy \((2t+1)\log_2 n+O(\log\log n)\).
	Their construction and ours share the use of multicoloring of the confusion graph to encode additional information through the choice of an assigned color.
	Specifically, their construction combines polynomial hashing with entropy recycling, using the choice among valid hash seeds to carry additional message information. 
	
	In a subsequent revision, Bui, Wang, Hagiwara, and Vu \cite{bui2026seed} introduced a second hashing stage and recycled seeds in both stages, improving the redundancy to \(2t\log_2 n+O_t(\log\log n)\).
	The revised result therefore matches the leading redundancy	term of our construction in the binary setting.
	However, the constructions differ in how the codeword suffixes are constructed: their construction uses two-stage hashing, whereas ours applies prime-based modular reduction. In addition, our construction applies directly to every fixed alphabet size \(q\ge2\), while their construction also allows correcting adjacent transposition errors.
	
	In another concurrent work, Gabrys, Singhvi, and Yaakobi \cite{gabrys2026making} constructed explicit binary \(t\)-edit-correcting codes with redundancy	\(2t\log_2 n+O_t(1)\).
	Their approach fixes a linear hash map and a target syndrome, then uses affine-system counting and a prefix-based encoder to map messages into a set of non-confusable words in the corresponding syndrome class.
	Encoding directly into this set avoids the overhead of appending and protecting message-dependent syndromes, yielding a sharper lower-order redundancy bound in the binary setting.
	Our construction instead provides a prime-based multicoloring approach that applies to arbitrary fixed
	alphabets, including those of non-prime-power size.

	\subsection{Technical overview}
	
	Our construction is guided by a simple decoding strategy.
	After at most \(t\) edits, the transmitted codeword lies among only \(O(n^t)\) length-\(n\) words within edit distance \(t\) of the received word. (See Lemma~\ref{lem:edit-degree}.) Thus, these candidates can be enumerated in polynomial time.
	Decoding therefore reduces to efficiently testing code membership and ensuring that at most one distinct candidate is valid.
	
	To ensure uniqueness, we combine graph coloring with edit-protected suffixes.
	Let \(m=n-\rho\), and consider the confusion graph on \(\cA^m\), whose edges join distinct prefixes at edit distance at most \(2t\).
	A (proper) coloring then assigns different colors to adjacent prefixes. Following the graph-coloring approach to code construction \cite{li-gabrys-farnoud2025}, we append to each prefix a length-\(\rho\) representation of its color, protected by a short \(t\)-edit-correcting code over \(\cA\).
	
	To see why this ensures uniqueness, suppose that two codewords are candidates for the same received word.
	Their edit distance is at most \(2t\) by the triangle inequality, and Lemma~\ref{lem:interval} gives the same distance bound for their corresponding prefixes and suffixes.
	Since distinct words in the short code have edit distance larger than \(2t\), the length-\(\rho\) suffixes must agree.
	They therefore represent the same color, which forces the prefixes to agree as well: distinct prefixes at distance at most \(2t\) cannot share a color. Thus, the two candidates coincide.
	
	The opportunity to reduce redundancy comes from allowing several colors for each prefix.
	Specifically, we replace the coloring with a \(J\)-fold multicoloring, assigning each prefix a set of \(J\) colors while requiring adjacent vertices to receive disjoint sets. The preceding uniqueness argument still applies, but each prefix now gives rise to \(J\) codewords.
	The code consequently has size \(Jq^m\) and redundancy \(\rho-\log_q J\): the choice among the \(J\) assigned colors carries additional message information.
	This multicoloring principle also appears in the concurrent	work \cite{bui2026binary}.
	The remaining challenge is to provide many colors per prefix without substantially increasing the length of their representations.
	
	We achieve this balance through a prime-based construction.
	First, the polynomial-evaluation method of Erd\H{o}s, Frankl, and F\"uredi
	\cite[Example~3.2]{erdos-frankl-furedi1985} gives an efficiently computable coloring \(f\) with \(O(\log n)\)-bit labels (Lemma~\ref{lem:proper-label}).
	We then use modular reduction, in the spirit of syndrome compression \cite{sima-syndrome2020}, to obtain several colors from each label.
	For a public list of \(L\) primes \(p_1,\ldots,p_L\), the color associated with index \(i\ge 0\) is
	$
	(i,f(x)\bmod p_{i+1}).
	$	
	We assign this pair to \(x\) only when the prime distinguishes \(f(x)\) from the labels of all its neighbors.
	More precisely, if \(y\) is adjacent to \(x\), we require
	\[
	f(x)\not\equiv f(y)\pmod{p_{i+1}}.
	\]
	This guarantees disjoint color sets: adjacent vertices using the same prime must have different residues.
	
	There are sufficiently many such primes because each nonzero difference between neighboring labels has only \(O(\log n)\) distinct prime factors.
	With \(O(n^{2t})\) neighbors per vertex, at most \(O(n^{2t}\log n)\) primes fail the required distinction.
	Choosing a sufficiently large number \(L=\Theta(n^{2t}\log n)\) of the smallest primes therefore leaves 
	\(J=\Theta(L)\) usable primes for every prefix (Lemma~\ref{lem:many-good}).
	Meanwhile, by the prime number theorem, we have \(p_L=O(L\log L)\), which keeps the color representations short. After protection by the short code, the suffix length is	\(\rho=4t\log_q n+O(\log\log n)\), whereas the color choice	carries \(\log_q J=2t\log_q n+O(\log\log n)\) message
	symbols. Hence, the net redundancy is
	\[
	\rho-\log_q J
	=2t\log_q n+O(\log\log n).
	\]
	
	Finally, the construction provides the efficient membership	test needed by the decoder.
	Evaluating \(f(x)\) uses only the values of \(x\) and its neighbors, so selecting or validating an assigned color requires information at graph distance at most two from
	\(x\). These neighborhoods and the public prime list can be enumerated in polynomial time.
	The short code has length \(O(\log n)\), so its construction and inversion also take polynomial time.
	A candidate can therefore be tested by inverting its suffix	and checking that the recovered color is assigned to its prefix. Together with candidate enumeration and uniqueness, this yields polynomial-time decoding, while color selection and suffix encoding give a polynomial-time encoder.

	\subsection{Discussion and open problems}
	
	For every fixed \(q\ge2\) and \(t\ge1\), the proposed \(t\)-edit-correcting codes have redundancy
	\(2t\log_q n+O(\log\log n)\) and deterministic polynomial-time encoding and decoding.
	Thus, it matches the leading redundancy term of the classical greedy bound.
	For substitution errors over prime-power alphabets, comparable or better redundancy is achievable by classical BCH codes, and the Gilbert-Varshamov argument admits polynomial-time derandomization for constant \(t\).
	Our result establishes that the greedy bound's leading term	is also efficiently achievable for arbitrary mixtures of insertions, deletions, and substitutions.
	
	We note that our encoder is not systematic. Although it preserves a length-\(m\) message prefix, the remaining message information is encoded through the choice of an	assigned color rather than appearing unchanged in the codeword.
	Obtaining a systematic encoder with comparable redundancy would be valuable, particularly for applications to document exchange protocols \cite{orlitsky1991,bernhard2019,cheng-jin-li-wu2022}.
	
	Beyond systematic encoding, a natural direction is to reduce the redundancy further.
	For \(t\ge2\), our bound remains above the recent existential bound \eqref{eq:rub} of \cite{en-gad2026}.
	An immediate goal is to attain that bound with deterministic polynomial-time encoding and decoding, ideally in systematic form.
	Within our construction, the prime-list size \(L\) can be reduced by bounding the number of distinct prime factors of the product of all neighboring label differences, rather	than bounding each difference separately.
	This refinement, however, improves the redundancy only by \(O(\log\log n)\) and leaves the leading coefficient unchanged.
	
	A more ambitious goal is to achieve redundancy \(t\log_q n+o(\log n)\), matching the leading term of the packing lower bound.
	For substitution errors, binary BCH codes already attain redundancy \(t\log_2 n+O(1)\) for every fixed \(t\).
	More recently, Yekhanin \cite{yekhanin2026nonlinear} constructed codes correcting two substitutions with redundancy
	\(2\log_q n+O(\sqrt{\log n})\) for every fixed \(q\ge2\), matching the packing bound in its leading term.
	Whether analogous leading-order optimality can be achieved by explicit edit-correcting codes for every fixed \(q\ge2\) and \(t\ge2\) remains open.
	
	\section{Preliminaries}\label{sec:prelim}
	
	Throughout, fix an alphabet \(\cA=\{0,\ldots,q-1\}\) with \(q\ge2\), and an edit budget \(t\ge1\). For $n\ge1$, let $\cA^n$ be the set of length-$n$ words over $\cA$. We use \(\cA^*\) to denote the set of finite words over $\cA$. For $x\in \cA^*$, we write \(|x|\) for its length, and \(x_{[\ell_0,\ell_1]}\) for its substring in positions \(\ell_0,\ldots,\ell_1\), where $\ell_0,\ell_1\geq 1$, i.e., the positions are one-based. 
	For finite words \(x,z\in\cA\), we write \(xz\) for their concatenation. 
	For an integer $M\geq 0$, define \(\nz{M}=\{0,\ldots,M-1\}\). 
	We use big-$O$ notation for asymptotic analysis, with subscripts indicating the parameters on which the implicit constants may depend (e.g., $O_{q,t}$). The notation $\widetilde O_{q,t}$ additionally suppresses polylogarithmic factors in the underlying asymptotic parameter. 
	
	We next introduce the edit distance between two finite words. To count changes when positions can shift, we place the two words in rows and allow gaps to record inserted or deleted symbols. 
	\begin{definition}[Edit alignment, cost, and distance]\label{def:alignment}
		Let \(\diamond\notin\cA\) denote a gap. An \emph{alignment} of \(x,y\in\cA^*\) is a sequence \(\mathsf A=((\widetilde x_i,\widetilde y_i))_{i=1}^{\ell}\) of \emph{columns} \( (\widetilde x_i,\widetilde y_i)\in (\cA\cup\{\diamond\})^2\setminus\{(\diamond,\diamond)\}\). The \emph{rows} \(\widetilde x,\widetilde y\) become \(x,y\), respectively, when gaps are removed. Columns with unequal symbols are of cost one, and any other column is a \emph{match} of cost zero. A column with a gap represents one insertion or deletion, and a column with unequal non-gap symbols represents one substitution. Define the \emph{alignment cost} and the \emph{edit distance}, respectively, by
		\[
		\operatorname{cost}(\mathsf A)=\bigl|\{i:\widetilde x_i\ne\widetilde y_i\}\bigr|,
		\qquad
		\distE(x,y)=\min_{\mathsf A\text{ aligning }x,y}\operatorname{cost}(\mathsf A).
		\]
	\end{definition}
	Equivalently, \(\distE(x,y)\) is the minimum length of an \emph{edit script}: an ordered sequence of insertions, deletions, and substitutions transforming \(x\) into \(y\), with each edit's position specified relative to the word immediately before the edit is applied. An edit alignment gives an edit script and vice versa.
	
	Although edit distance can increase under arbitrary coordinate restrictions, the following lemma shows that it does not increase when two equal-length words are restricted to the same coordinate interval; see also \cite[Lemma~18]{xie-chen2026}. For completeness, the proof of Lemma~\ref{lem:interval} is given in Appendix~\ref{app:interval}.
	
	\begin{lemma}[Edit distance under interval restriction]\label{lem:interval}
		Let \(x,y\in\cA^n\), and let \(I=\{\ell_0,\ell_0+1,\ldots,\ell_1\}\), where \(1\le\ell_0\le\ell_1\le n\).
		Write \(x_I=x_{[\ell_0,\ell_1]}\) and \(y_I=y_{[\ell_0,\ell_1]}\). Then
		\[
		\distE(x_I,y_I)\le\distE(x,y).
		\]
	\end{lemma}

	For a finite word $x\in \cA^*$, its \emph{edit ball} of radius $t$ is defined to be
	\[
	\cB_t(x)=\{y\in\cA^*: \distE(x,y)\le t\}.
	\]
	Since the $q$-ary edit channel permits any total of at most $t$ insertions, deletions, and substitutions, the edit ball $\cB_t(x)$ is the set of all possible received words if $x$ is transmitted. 
	The following lemma gives a standard upper bound on the size of an edit ball, whose proof is given in Appendix~\ref{app:edit-degree}.
	
	\begin{lemma}[Volume of an edit ball]\label{lem:edit-degree}
		For every \(m\ge1\) and \(x\in\cA^m\), the number of edit scripts of length at most \(t\) starting
		from \(x\), including the empty script, is at most
		$
		\sum_{i=0}^{t}(2q(m+t+1))^i.
		$
		Consequently, defining
		\begin{align}
			r_t(m):=(t+1)\bigl(2q(m+t+1)\bigr)^t,\label{eq:r-bound}
		\end{align}
		we have
		\[
		|\cB_t(x)|\le r_t(m)=\Theta_{q,t}(m^t).
		\]
	\end{lemma}

	A subset \(\cC\subseteq\cA^n\) is said to be a \emph{$t$-edit-correcting} code if the edit balls of radius $t$ are disjoint for distinct codewords in $\cC$. The redundancy of $\cC$ is defined to be \(n-\log_q|\cC|\), in \(q\)-ary symbols, and the edit distance of $\cC$ is defined to be 
	\[
	\distE(\cC)=\min_{c,c'\in\cC,c\neq c'}\distE(c,c').
	\]
	
	To construct a $t$-edit-correcting code, one needs to identify pairs of words whose edit balls do not intersect. 
	For \(x,x'\in\cA^n\) with $x\neq x'$,  the balls $\cB_t(x)$ and $\cB_t(x')$ intersect if and only if \(\distE(x,x')\le2t\). Thus, a code is {$t$-edit-correcting} if and only if its edit distance is greater than $2t$. 
	
	Recall that the problem of constructing large codes is equivalent to finding a large {independent set} in the so-called confusion graph, defined as follows.
	\begin{definition}[Confusion graph]\label{def:graph}
		For each \(m\ge 1\), the \emph{confusion graph} \(\cG_m\) is the finite simple graph with vertex set \(\cA^m\) in which two distinct vertices \(x,x'\) are adjacent if and only if
		\[
		\distE(x,x') \le 2t.
		\]
		Adjacent vertices are said to be \emph{confusable}.
	\end{definition}
	
	By Definition~\ref{def:graph}, a code is $t$-edit-correcting if and only if it is an independent set in the confusion graph. Moreover, a coloring of the confusion graph corresponds to a partition of the vertex set into independent subsets, and thus each {color class} is a $t$-edit-correcting code.
	For our subsequent discussion, it is useful to introduce \emph{multicoloring} (or \emph{fractional coloring}), a generalization of graph coloring in which each vertex is assigned a set of colors.
	
	\subsection{Graph multicoloring}\label{sec:multi}
	
	Let \(\cG=(V,E)\) be a finite simple graph. The set of vertices adjacent to $v\in V$ is denoted by \(\cN(v):=\{u\in V: \{u,v\}\in E\}\). Let $\cP$ be a finite set of available colors. We refer to $\cP$ as the \emph{palette}. For an integer $J\ge1$, an \emph{allocation} of size \(J\) from the palette \(\cP\) assigns each vertex \(v\in V\) a set $S(v)\subseteq\cP$ of distinct colors satisfying 
	\[
	|S(v)|=J,\qquad S(u)\cap S(v)=\emptyset\quad\text{for every }u\in \cN(v).
	\]
	The sets \(S(v),v\in V\) form a $J$-fold multicoloring. 
	
	A multicoloring can be constructed explicitly using polynomial evaluation \cite[Example~3.2]{erdos-frankl-furedi1985}, \cite[Section~4.3]{kuhn2009}; see also \cite{li-gabrys-farnoud2025,ye-ge-substring2026} for recent	applications of this method to the construction of edit-correcting codes.
	
	\subsubsection{A coloring by polynomial evaluation}\label{subsec:poly-color}
	
	Fix an injective \emph{identifier map} \(\id:V\to\nz{2^K}\), where \(K\ge1\) is an integer. Let \(r\ge 1\) be an integer upper bound on the maximum degree of $\cG$, i.e., $r\geq \max_{v\in V}|\cN(v)|$. 
	Choose
	\begin{equation*}\label{eq:field}
		Q=2^{\ceil{\log_2(rK+1)}}.
	\end{equation*}
	Thus, \(rK<Q<2(rK+1)\). 
	
	Let \(\F_Q\) be the finite field of order \(Q\),\footnote{The field $\F_Q$ is taken to have characteristic two for simplicity. Any other finite field of order greater than $rK$ also works; for example, one can also take $Q$ to be the least prime greater than $rK$.} and fix a bijection between \(\F_Q\) and \(\nz{Q}\). In what follows, we identify each field element with its corresponding integer under this bijection whenever the intended interpretation is clear from context.
	We order the field elements by their corresponding integers.
	
	\begin{definition}[Identifier polynomial and separating point]\label{def:separating-point}
		For a vertex \(v\in V\), write the base-\(Q\) representation of its identifier $\id(v)$ as \(v_0,\ldots,v_{K-1}\in \nz{Q}\), retaining leading zeros:
		\[
		\id(v)=\sum_{e=0}^{K-1}v_eQ^e.
		\] 
		Interpreting each $v_e$ as an element in $\F_Q$, define the \emph{identifier polynomial} of $v$ to be \[
		g_v(X)=\sum_{e=0}^{K-1}v_eX^e\in\F_Q[X].
		\]
		A point \(\beta\in\F_Q\) \emph{separates} \(v\) if \(g_v(\beta)\ne g_u(\beta)\) for every \(u\in\cN(v)\). Let \(\alpha_v\) be the first separating point in $\F_Q$ with the ordering induced by bijection between $\F_Q$ and $\nz{Q}$.
	\end{definition}
	
	The following lemma gives a coloring based on separating points of $v\in V$ and evaluation of their identifier polynomial.
	
	\begin{lemma}[Coloring]\label{lem:proper-label}
		For each $v\in V$, there exists $\alpha_v\in\F_Q$ that separates $v$. Moreover, the map $f\colon V\to \nz{Q^2}$ defined by
		\begin{align*}
			f(v)=Q\alpha_v+g_v(\alpha_v)\label{eq:color}
		\end{align*}
		gives a coloring, where $g_v\in\F_Q[X]$ is the identifier polynomial of $v$ and $f(v)$ is computed by viewing \(\alpha_v, g_v(\alpha_v)\in \F_Q\) as integers in $\nz{Q}$ under the fixed bijection between $\F_Q$ and $\nz{Q}$.
	\end{lemma}

	\begin{proof}
		We first prove the existence of \(\alpha_v\). The identifier map is injective, so \(g_v-g_u\) is a nonzero polynomial of degree less than \(K\) for each \(u\in \cN(v)\). Each such polynomial has at most \(K-1\) roots. Therefore, at most \(|\cN(v)|(K-1)\le r(K-1)<Q\) field elements do not separate $v$, proving existence. 
		
		We next show that the values $f(v),v\in V$ form a coloring. If $u\in\cN(v)$ and $f(u)=f(v)$, then $(\alpha_u,g_u(\alpha_u))=(\alpha_v,g_v(\alpha_v))$. In other words, the base-\(Q\) representations of $f(u)$ and $f(v)$ give the same separating point and the same evaluation, contradicting the definition of separating points (Definition~\ref{def:separating-point}). 
	\end{proof}
	
	\begin{remark}[Complexity]\label{rem:color-time}
		Assume that $|V|=2^{O(n)}$ and $K,r$ are polynomial in $n$, and that the identifier $\id(v)$ is computable in polynomial time in $n$. Then the color evaluation $f(v)$ takes polynomial time in $n$. Moreover, assuming \(\cN(v)\) can be efficiently enumerated, the search of separating points for $v$ examines only polynomially many field elements and identifier polynomials of its neighbors. See Appendix~\ref{app:graph-costs} for a detailed bit-cost analysis.
	\end{remark}
	
	\subsubsection{A multicoloring and enumerative coding}
	
	Now let \(f:V\to\nz{2^B}\) be a coloring of $\cG$, where \(B\ge1\) is an integer. Thus, \(f\) assign a $B$-bit label to each vertex in \(V\). To extend $f$ to a multicoloring, we adapt an approach remarked in the full version of \cite{kuhn2009}. 
	
	Specifically, let \(p_i\) denote the \(i\)-th smallest prime. The idea suggested in \cite[p.~11 of the full version]{kuhn2009} is to assign each vertex $v$ the color set formed by pairs $(i,\id(v)\bmod p_i)$, using only primes $p_i$ for which $\id(u)\not\equiv \id(v) \pmod{p_i}$ for all $u\in \cN(v)$. The Chinese remainder theorem bounds the number of primes that fail to distinguish a given pair of identifiers, while the prime number theorem controls the sizes of the candidate primes.

	Since $f(v)$ is a coloring, $f(u)\ne f(v)$ for every $u\in\cN(v)$. Thus, $f(v)$ already distinguishes $v$ from its neighbors and can replace $\id(v)$ in this construction. Therefore, we can construct a $J$-fold multicoloring by assigning each vertex pairs of the form $(i, f(v)\bmod p_i)$ where $p_i$ is the prime such that $f(u)\not\equiv f(v) \pmod{p_i}$ for all $u\in \cN(v)$. The resulting color sets are then disjoint for adjacent vertices. Hence, to obtain a $J$-fold multicoloring, it suffices to find at least $J$ such primes for each vertex.
	In the following, we restrict attention to the first $L$ primes with
	\begin{equation}\label{eq:L}
		L=4(r+1)(B+1), 
	\end{equation} where $r$ is the maximum degree of $\cG$. 
	To estimate $p_L$, we use the standard bounds 
	\begin{equation}\label{eq:prime-bound}
		p_L\ge L,\qquad p_L=O(L\log L),
	\end{equation} where the second bound follows from the prime number theorem. 
	The next definition formalizes the required property for the multicoloring construction.
	\begin{definition}[Good primes]\label{def:good}
		For $v\in V$ and \(i\in\nz{L}\), the index $i$ and the corresponding prime \(p_{i+1}\) are \emph{good} for \(v\) if
		\[
		f(v)\not\equiv f(u)\pmod{p_{i+1}}
		\quad\text{for every }u\in\cN(v).
		\]
		Let \(\good(v)\) be the set of primes good for \(v\), ordered increasingly. For \(0\le j<|\good(v)|\), let
		\(i_v(j)\in\nz{L}\) be the unique index such that \(p_{i_v(j)+1}\) is the \((j+1)\)-st smallest element
		of \(\good(v)\).
	\end{definition}
	
	We will only use the first $J\leq L/2$ good primes in $\good(v)$ for each vertex $v$, which are among the first $L$ smallest primes, to construct the $J$-fold multicoloring. More precisely, let
	\begin{equation}\label{eq:J}
		J=q^{\floor{\log_q(L/2)}}.
	\end{equation}
	With the palette \(\cP:=\bigcup_{i\in\nz{L}}\{i\}\times\nz{p_{i+1}}\), assign to each vertex \(v\) the color set
	\begin{equation}\label{eq:multi-color}
		S(v):=\{\pi_v(j):=\bigl(i_v(j),f(v)\bmod p_{i_v(j)+1}\bigr)\mid j\in\nz{J}\}\subseteq\cP.
	\end{equation}
	To distinguish the (scalar) color \(f(v)\) with the color \(\pi_v(j)\), we also call \(\pi_v(j)\) a color pair.
	
	\begin{lemma}[Multicoloring]\label{lem:many-good}
		The sets $S(v),v\in V$ in \eqref{eq:multi-color} form a $J$-fold multicoloring.
	\end{lemma}
	
	\begin{proof}
		We first show that $|S(v)|=J$ for all $v\in V$. In other words, we show that the colors $\pi_v(j)=(i_v(j),f(v)\bmod p_{i_v(j)+1}),j\in\nz{J}$ are indeed distinct for each vertex $v\in V$.
		
		For each vertex \(u\in \cN(v)\), we have 
		$
		0<\abs{f(v)-f(u)}<2^B.
		$ Since the product of \(B\ge 1\) distinct primes is at least \(2^B\), the nonzero integer $\abs{f(v)-f(u)}$ has fewer than \(B\) distinct prime divisors.
		Thus, each neighbor $u\in\cN(v)$ excludes fewer than $B$ primes from $\good(v)$,\footnote{The same conclusion can be obtained by an application of the Chinese remainder theorem.} and the union of excluded primes has size at most \(|\cN(v)|B\le rB<L/4\), where the last inequality follows from \eqref{eq:L}. Hence, there are more than \(3L/4\) good primes for each vertex $v$, i.e., $|\good(v)|>3L/4$. By \eqref{eq:J}, we have \(J\le L/2\) and thus the $J$ colors $\pi_v(j),j\in\nz{J}$ are well defined and all distinct.
		
		It remains to prove that \(S(u)\cap S(v)=\emptyset\) for \(u\in\cN(v)\). Suppose that a color belongs to both \(S(u)\) and \(S(v)\), and the first coordinate of the color is the index \(i\). Then the second coordinate of the color gives \(f(u)\equiv f(v)\pmod{p_{i+1}}\). But \(i\) is good for \(v\), and \(u\) is a neighbor of $v$, contradicting the definition of good primes (Definition~\ref{def:good}).
	\end{proof}

	Note that if \(f\) is a coloring of the confusion graph \(\cG_m\) on \(\cA^m\), then each color class is a \(t\)-edit-correcting code, and \(f(x)\) is called the \emph{syndrome} of \(x\in\cA^m\). In this setting, syndrome compression \cite{sima-syndrome2020} replaces \(f(x)\) with the pair \((b,f(x)\bmod b)\), where \(b\) is a positive integer chosen so that \(f(x)\not\equiv f(y)\pmod{b}\) for every \(y\in\cN(x)\). The idea of choosing good primes as moduli for this purpose also appears in	\cite{con-li2025} and the recent work \cite{ye-ge-substring2026}.

	\paragraph{Enumerative coding.}
	Fix \(v\in V\), and order \(S(v)=\{\pi_v(j):j\in\nz{J}\}\) by the first coordinate of its colors. Since \(i_v(j)\) is strictly increasing in \(j\), the map \(j\mapsto\pi_v(j)\) is an order-preserving bijection from \(\nz{J}\) onto \(S(v)\). This yields an enumerative coding \cite{cover1973} of \(S(v)\), with explicit ranking and unranking procedures as follows.
	
	Given a zero-based rank \(j\in\nz{J}\), select the \((j+1)\)-st	smallest prime in \(\good(v)\), namely, \(p_{i_v(j)+1}\), and return the corresponding color
	\[
	\pi_v(j)=\bigl(i_v(j),f(v)\bmod p_{i_v(j)+1}\bigr).
	\]
	Conversely, given a color \((i,a)\in S(v)\), recover its zero-based	rank by counting the primes in \(\good(v)\) smaller than \(p_{i+1}\):
	\[
	j=\bigl|\{p\in\good(v):p<p_{i+1}\}\bigr|.
	\]
	Both operations can be done by scanning the first \(L\) primes and testing which are good for \(v\).
	
	\begin{remark}[Complexity]\label{rem:allocation-time}
		Assume that \(|V|=2^{O(n)}\) and \(B,r\) are polynomial in \(n\), that \(f\) can be evaluated in polynomial time, and that \(\cN(v)\) can be efficiently enumerated for each \(v\in V\). Then generating the first \(L\) primes and performing each ranking or unranking operation take polynomial time in \(n\). See Appendix~\ref{app:graph-costs} for details.
	\end{remark}
	
	\section{The edit code}\label{sec:edit}
	
	In this section, we construct a \(t\)-edit-correcting code \(\cC\subseteq\cA^n\) with redundancy \(2t\log_q n+O_{q,t}(\log\log n)\).
	The construction combines \(J\)-fold multicoloring on the confusion graph $\cG_m$ with a short edit-correcting code over \(\cA\).
	For each vertex \(x\in\cA^m\), we convert each of its \(J\)	colors into a \(q\)-ary word and protect the resulting word using the short code. Appending these protected color representations to \(x\) produces \(J\) full codewords. Taking these codewords over all \(x\in\cA^m\) yields our edit code of size \(Jq^m\) and length \(n\).
	
	We choose all auxiliary parameters in terms of \(n\), so that the multicoloring applies to every length \(m\le n\).	The actual choice of \(m\) is then determined by the length of the protected color representation.
	
	\paragraph{Setting up the parameters.}
	
	Recall that two distinct vertices of \(\cG_m\) are adjacent	when their edit distance is at most \(2t\). For every \(1\le m\le n\), Lemma~\ref{lem:edit-degree} and the monotonicity of \(r_{2t}\) (see Eq.~\eqref{eq:r-bound}) give
	\[
	\max_{x\in\cA^m}|\cN(x)|
	\le \max_{x\in\cA^m}|\cB_{2t}(x)|
	\le r_{2t}(m)
	\le r_{2t}(n).
	\]
	We therefore choose the integer maximum-degree bound
	\begin{equation}\label{eq:r}
		r=r_{2t}(n)
		=\Theta_{q,t}(n^{2t}),
	\end{equation}
	which applies uniformly to all prefix lengths \(1\le m\le n\).
	Following Section~\ref{sec:multi}, set
	\begin{gather*}
		K=n\ceil{\log_2 q},\qquad
		Q=2^{\ceil{\log_2(rK+1)}},\qquad
		B=2\ceil{\log_2(rK+1)},\\
		L=4(r+1)(B+1),\qquad
		J=q^{\floor{\log_q(L/2)}}.
	\end{gather*}
	In particular, \(J\) is a power of \(q\) and satisfies
	\begin{align}
		\frac{L}{2q}<J\le\frac{L}{2}.\label{eq:J-bound}
	\end{align}
	These choices give  
	\begin{align}
		L=\Theta_{q,t}(r\log n).\label{eq:L-a}
	\end{align}
	The bounds on the \(L\)-th prime in \eqref{eq:prime-bound} therefore imply
	\begin{equation}\label{eq:edit-prime-sizes}
		L\le p_L=O_{q,t}\bigl(r(\log n)^2\bigr).
	\end{equation}
	
	\paragraph{Multicoloring the confusion graph.}
	
	For any \(1\le m\le n\) and \(x=x_1\cdots x_m\in\cA^m\), define
	\[
	\id(x):=\sum_{\mu=1}^m x_\mu q^{m-\mu}.
	\]
	This identifier map is injective on \(\cA^m\) and satisfies
	$
	0\le\id(x)<q^m\le q^n\le 2^K.
	$
	As in Section~\ref{subsec:poly-color}, fix a bijection between \(\F_Q\) and \(\nz{Q}\). From the base-\(Q\) representation of \(\id(x)\), form the identifier polynomial \(g_x\in\F_Q[X]\) as in Definition~\ref{def:separating-point}.
	
	Let \(\alpha_x\) be the first separating point for \(x\). By Lemma~\ref{lem:proper-label}, such a point exists, and
	\begin{equation}\label{eq:prefix-color}
		f(x)=Q\alpha_x+g_x(\alpha_x)
	\end{equation}
	defines a coloring \(f:\cA^m\to\nz{Q^2}=\nz{2^B}\) of \(\cG_m\). Here, the field elements \(\alpha_x,g_x(\alpha_x)\in\F_Q\) in \eqref{eq:prefix-color} are interpreted as their integer representations in \(\nz{Q}\) under the fixed bijection.
	
	Applying Lemma~\ref{lem:many-good} to this coloring yields a \(J\)-fold multicoloring \(S(x),x\in\cA^m\), where
	\[
	S(x)=\{\pi_x(j):j\in\nz{J}\},\qquad
	\pi_x(j)=
	\bigl(i_x(j),f(x)\bmod p_{i_x(j)+1}\bigr).
	\]
	
	\subsection{Encoding and redundancy}\label{sec:serialization}
	
	We next convert each color pair into a fixed-length \(q\)-ary representation and protect this representation against edits. Recall that the palette is
	\[
	\cP=\bigcup_{i\in\nz{L}}\{i\}\times\nz{p_{i+1}}.
	\]
	For an integer \(\ell\ge1\) and \(N\in\nz{q^\ell}\), let \(\rep_\ell(N)\) denote the length-\(\ell\) base-\(q\) representation of \(N\), including leading zeros. Set
	\[
	h_1=\ceil{\log_q L},\qquad
	h_2=\ceil{\log_q p_L},\qquad
	h=h_1+h_2.
	\]
	Define the \emph{serialization} map on the palette:
	\begin{equation*}\label{eq:serialization}
		\ser:\cP\longrightarrow\cA^h,\qquad
		\ser(i,\eta)=\rep_{h_1}(i)\rep_{h_2}(\eta).
	\end{equation*}
	The map \(\ser\) is injective because the two coordinates have unique representations of fixed lengths.
	By \eqref{eq:r}, \eqref{eq:edit-prime-sizes}, and the bound \(L=\Theta_{q,t}(r\log n)\), its output length satisfies
	\begin{equation}\label{eq:edit-color-length}
		h=4t\log_q n+O_{q,t}(\log\log n).
	\end{equation}
	
	To protect the serialized color pairs, we use the following short edit-correcting code, for which a standard greedy construction suffices. The proof of Lemma~\ref{lem:short-edit} is given in Appendix~\ref{app:short-edit}.
	
	\begin{lemma}[Short edit code]\label{lem:short-edit}
		Fix integers \(q\ge2\) and \(t\ge1\).
		For every integer \(h\ge1\), the minimum 
		\[
		\rho:=
		\min\bigl\{\ell\in\mathbb{Z}:
		\ell\ge h,\ q^\ell\ge q^h r_{2t}(\ell)\bigr\}
		\]
		is well defined and satisfies
		$
		\rho=h+O_{q,t}(\log(h+1)).
		$
		Moreover, there is an injective encoder
		\[
		E:\cA^h\longrightarrow\cA^{\rho}
		\]
		whose image is a \(t\)-edit-correcting code.
	\end{lemma}
	
	\begin{remark}[Complexity of the short code]\label{rem:short-time}
		For fixed \(q\) and \(t\), the greedy construction of the short code, including its encoder table, takes \(O_{q,t}(q^h(h+1)^{2t+2})\) time.
		By \eqref{eq:edit-color-length}, this is \(\widetilde O_{q,t}(n^{4t})\). This construction is performed once as part of public setup. See Appendix~\ref{app:short-costs} for details.
	\end{remark}
	
	Applying Lemma~\ref{lem:short-edit} with the length \(h\) in \eqref{eq:edit-color-length} gives \(\rho=O_{q,t}(\log n)\).
	We can now fix the prefix length $m$ and message length $k$ as follows:
	\begin{equation}\label{eq:edit-lengths}
		m=n-\rho,\qquad k=m+\log_q J.
	\end{equation}
	Since \(J\) is a power of \(q\), it follows that \(k\) is an integer. Moreover, \(m\ge1\) for all sufficiently large \(n\).
	We are now ready to describe the code construction.
	
	\begin{construction}\label{con:edit}
		Identify each pair \((x,j)\in\cA^m\times\nz{J}\) with a message \(x\rep_{k-m}(j)\in\cA^k\). Define the protected suffix
		$
		z=E(\ser(\pi_x(j)))\in\cA^{\rho},
		$
		and define the encoder \(C:\cA^k\to\cA^n\) by
			$C(x,j)=xz.$
		Let \(\cC\) be the image of \(C\).
	\end{construction}
	
	\begin{lemma}[Redundancy]
		\label{lem:edit-redundancy}
		The encoder \(C\) is injective. Consequently, \(|\cC|=Jq^m=q^k\), and the redundancy of \(\cC\) is
		\[
		n-k=2t\log_q n+O_{q,t}(\log\log n).
		\]
	\end{lemma}
	
	\begin{proof}
		Suppose that \(C(x,j)=C(x',j')\). Since both prefixes \(x\) and \(x'\) have length \(m\), we must have \(x=x'\).
		The injectivity of \(E\) and \(\ser\) then gives \(\pi_x(j)=\pi_x(j')\). Since the map \(j\mapsto\pi_x(j)\) is injective, it follows that \(j=j'\). Thus, \(C\) is injective.
		
		To calculate the redundancy, note that from \eqref{eq:J-bound} and \eqref{eq:L-a} we have \(J=\Theta_q(L)\) and \(L=\Theta_{q,t}(n^{2t}\log n)\), so
		$
		\log_q J
		=2t\log_q n+O_{q,t}(\log\log n).
		$
		Using \eqref{eq:edit-lengths}, Lemma~\ref{lem:short-edit}, and \eqref{eq:edit-color-length}, we obtain
		\begin{align*}
			n-k
			&=\rho-\log_q J
			\\
			&=h+O_{q,t}(\log(h+1))-\log_q J
			\\
			&=2t\log_q n+O_{q,t}(\log\log n).
		\end{align*}
	\end{proof}
	
	\begin{remark}[Encoding time]\label{rem:encoder-time}
		The encoder can be implemented in \(\widetilde O_{q,t}(n^{6t+2})\) time.
		The dominant cost is selecting the color pair \(\pi_x(j)\) assigned to the prefix \(x\).
		
		To compute \(f(x)\), we search for its first separating element among the \(Q=O_{q,t}(rn)\) elements of \(\F_Q\). At each field element, Horner's rule evaluates the identifier polynomials of \(x\) and its neighbors: at most \(r+1\) polynomials, each of degree at most
		\(K-1\), where \(K=O_q(n)\).
		Thus, one evaluation of \(f\) takes $\widetilde O_{q,t}\bigl(Q(r+1)K\bigr)=\widetilde O_{q,t}(r^2n^2)$ time. This bound also absorbs the costs of neighborhood
		enumeration and identifier formation.
		
		Selecting \(\pi_x(j)\) requires \(f(x)\) and \(f(u)\) for every \(u\in\cN(x)\), hence at most \(r+1\) evaluations
		of \(f\). The subsequent scan of the public list of the first $L$ primes, serialization, and short-code encoding contribute only	lower-order terms. Since \(r=\Theta_{q,t}(n^{2t})\), the overall encoding cost is $\widetilde O_{q,t}\bigl((r+1)r^2n^2\bigr)=\widetilde O_{q,t}(n^{6t+2})$.
		See Appendix~\ref{app:composition} for details.
	\end{remark}
	
	\subsection{Decoding and correctness}
	
	Our decoder enumerates all length-\(n\) words within edit distance \(t\) of the received word and tests whether they belong to \(\cC\). If at most \(t\) edits have occurred, the transmitted codeword is among these candidates. Moreover, by Lemma~\ref{lem:edit-degree}, the number of candidates is \(O_{q,t}(n^t)\), and thus enumerating all the candidates takes time polynomial in $n$ for constant $q$ and $t$.
	
	To validate a candidate, we split it into a length-\(m\) prefix	and a length-\(\rho\) suffix, and check that the suffix encodes a color assigned to the prefix. We then show that at	most one distinct candidate can pass these tests.
	
	\paragraph{Decoding procedure.}
	The decoder is defined for every received word \(y\in\cA^*\).
	If \(||y|-n|>t\), reject \(y\). Otherwise, enumerate all edit scripts of length at most \(t\) starting from \(y\), and process each length-\(n\) word \(\hat c\) they produce.
	Every such candidate satisfies \(\distE(\hat c,y)\le t\).
	
	Write \(\hat c=\hat x\hat z\), where \(\hat x\in\cA^m\) and	\(\hat z\in\cA^{\rho}\). Apply the following tests,
	discarding the candidate if any test fails.
	\begin{enumerate}
		\item \emph{Suffix validation.}
		Verify that \(\hat z\in E(\cA^h)\), and recover the unique	word \(s\in\cA^h\) satisfying \(E(s)=\hat z\).
		
		\item \emph{Deserialization.}
		Split \(s\) into blocks of lengths \(h_1\) and \(h_2\),	and let \(i\) and \(\eta\) be their respective base-\(q\) values. Require \(i<L\) and \(\eta<p_{i+1}\), so that \((i,\eta)\in\cP\) and \(s=\ser(i,\eta)\).
		
		\item \emph{Color validation and rank recovery.}
		Compute \(f(\hat x)\) and \(f(u)\) for every \(u\in\cN(\hat x)\). Require that \(i\) is good for \(\hat x\) and that 
		\[
		\eta=f(\hat x)\bmod p_{i+1}.
		\]
		Determine the zero-based rank of \(p_{i+1}\) in	\(\good(\hat x)\):
		\[
		j=\bigl|\{p\in\good(\hat x):p<p_{i+1}\}\bigr|.
		\]
		Finally, require \(j<J\). These conditions ensure that \((i,\eta)=\pi_{\hat x}(j)\in S(\hat x)\).
	\end{enumerate}
	As soon as a candidate passes all tests, return the message	corresponding to \((\hat x,j)\), namely, \(\hat x\,\rep_{k-m}(j)\).
	If no candidate passes, reject \(y\).
	
	\begin{lemma}[Decoder correctness]\label{lem:decoder-correctness}
		The decoder recovers the transmitted message whenever at most \(t\) edits occur and rejects every received word	outside all radius-\(t\) edit balls around codewords.
	\end{lemma}
	
	\begin{proof}
		First, every passing candidate \(\hat c=\hat x\hat z\) satisfies
		$
		\hat z=E(\ser(\pi_{\hat x}(j)))
		$
		for the recovered rank \(j\). Hence, we have \(\hat c=C(\hat x,j)\in\cC\). Since \(\hat c\) is generated by an edit script of length at most \(t\) starting from \(y\), we also have \(\distE(\hat c,y)\le t\).
		Thus, the decoder rejects every word outside all radius-\(t\) edit balls around codewords.
		
		Conversely, suppose that \(y\) is obtained from a transmitted codeword \(C(x,j)\) by at most \(t\) edits. It follows that \(||y|-n|\le t\) and the initial length test passes. Since each edit is reversible, the enumeration includes an edit script producing \(C(x,j)\) from
		\(y\). This candidate passes all validation tests, so at least one passing candidate exists.
		
		It remains to prove uniqueness. Suppose that \(\hat c=\hat x\hat z\) and
		\(\hat c'=\hat x'\hat z'\) both pass. By the triangle inequality,
		\[
		\distE(\hat c,\hat c')
		\le \distE(\hat c,y)+\distE(y,\hat c')
		\le 2t.
		\]
		Since both candidates have the same prescribed prefix and suffix lengths, Lemma~\ref{lem:interval} gives
		\begin{align}
			\distE(\hat x,\hat x')\le2t,
			\qquad
			\distE(\hat z,\hat z')\le2t.\label{eq:affix-bounds}
		\end{align}
		Both suffixes belong to the short code \(E(\cA^h)\), whose distinct codewords have edit distance at least \(2t+1\). Therefore, \(\hat z=\hat z'\). By the injectivity of \(E\) and \(\ser\), the two suffixes
		represent the same color.
		
		If \(\hat x\ne\hat x'\), the prefix distance bound in \eqref{eq:affix-bounds} implies that \(\hat x\) and \(\hat x'\) are adjacent in \(\cG_m\).
		Their color sets must then be disjoint by Lemma~\ref{lem:many-good}, contradicting the
		fact that the two candidates use the same color.
		Thus, \(\hat x=\hat x'\), and consequently \(\hat c=\hat c'\). The injectivity of \(C\) now implies	that the recovered messages are also equal.
		
		Hence, at most one distinct candidate passes the tests.
		When at most \(t\) edits occur, this candidate is the transmitted codeword, so returning the first passing candidate recovers the original message.
	\end{proof}
	
	\begin{remark}[Decoding time]\label{rem:decoder-time}
		The decoder runs in \(\widetilde O_{q,t}(n^{7t+2})\) time, with the dominant cost being candidate validation. 
		
		For a received word passing the initial length test, Lemma~\ref{lem:edit-degree} bounds the number of generated candidates by \(O_{q,t}(n^t)\).
		Validating a candidate uses the same color evaluations and prime scan as selecting an assigned color pair in the encoding process, together with suffix inversion and deserialization.
		Thus, each validation takes	\(\widetilde O_{q,t}(n^{6t+2})\) time, as in Remark~\ref{rem:encoder-time}. The total validation cost is therefore
		$O_{q,t}(n^t)\widetilde O_{q,t}(n^{6t+2})=\widetilde O_{q,t}(n^{7t+2})$.
		See Appendix~\ref{app:composition} for details.
	\end{remark}
	
	Lemmas~\ref{lem:edit-redundancy} and~\ref{lem:decoder-correctness},	together with the complexity analysis in Appendix~\ref{app:runtime}, establish Theorem~\ref{thm:edit-main}.
	
	\section*{Acknowledgment of AI use}
	The underlying research ideas and proofs were proposed by GPT-5.6 Sol while H.~Xie attempted to extend the work \cite{xie-chen2026} to correcting a larger number of edits in the nanopore channel. The first draft of the results presented in this manuscript was produced by OpenAI Codex on September 7, 2026. The authors verified and substantially reworked the draft, and assume full responsibility for the content of this manuscript.
	
	\bibliographystyle{alphaurl}
	\bibliography{references}
	
	\appendix
	\section{Proof of Lemma~\ref{lem:interval}}\label{app:interval}
		Fix a minimum-cost alignment of \(x\) and \(y\). Since \(|x|=|y|\), the two rows of the alignment contain the same number \(\delta\) of gaps. Let \(\sigma\) be the number of mismatched pairs of non-gap symbols. Then
		\[
		\distE(x,y)=2\delta+\sigma.
		\]
		
		If \(|I|\le\delta\), we can transform \(x_I\) into \(y_I\) using at most \(|I|\) substitutions, so
		\[
		\distE(x_I,y_I)\le |I|\le\delta\le 2\delta+\sigma=\distE(x,y).
		\]
		Next, we assume that \(|I|>\delta\).
		Exactly \(\delta\) symbols of \(x\) are aligned with gaps in the \(y\)-row, so the alignment contains exactly \(n-\delta\) columns with two non-gap symbols. List these columns in alignment order, and let \((i_\nu,j_\nu)\) be the positions of their symbols in
		\(x\) and \(y\), respectively.
		Before the \(\nu\)-th such column, each row contains \(\nu-1\) symbols paired with non-gap symbols and at most \(\delta\) symbols paired with gaps. Hence, for \(1\le\nu\le n-\delta\) we have
		\[
		\nu\le i_\nu,j_\nu\le\nu+\delta.
		\]
		
		Retain the pairs with indices
		\[
		\nu=\ell_0,\ell_0+1,\ldots,\ell_1-\delta.
		\]
		Since \(|I|>\delta\) and \(\ell_1\le n\), this range is nonempty and contained in \(\{1,\ldots,n-\delta\}\). For every retained pair, the preceding bounds give
		\[
		\ell_0\le i_\nu,j_\nu\le\ell_1.
		\]
		Thus, both symbols of each retained pair lie in the words \(x_I\) and \(y_I\).
		
		Denote the resulting subsequences by \(\hat{x}\) and \(\hat{y}\), respectively. Both have length \(|I|-\delta\), and at most	\(\sigma\) of their corresponding symbols differ. We can therefore transform \(x_I\) into \(y_I\) by deleting the \(\delta\) symbols not retained in \(\hat{x}\), making at most \(\sigma\) substitutions to obtain \(\hat{y}\), and inserting the \(\delta\) missing symbols to recover \(y_I\). Consequently, \(\distE(x_I,y_I)\le 2\delta+\sigma=\distE(x,y)\).
	
	\section{Proof of Lemma~\ref{lem:edit-degree}}\label{app:edit-degree}
		An edit script is an ordered sequence of edit operations. During any script of length at most \(t\) starting from \(x\), every intermediate word has length at most \(m+t\). For a word of length \(\ell\), there are \(q(\ell+1)\) possible insertions, \((q-1)\ell\) possible substitutions, and \(\ell\) possible deletions. Thus, at each step, the number of possible edits is at most
		\[
		q(\ell+1)+(q-1)\ell+\ell
		=2q\ell+q
		\le 2q(m+t+1).
		\]
		It follows that the number of scripts of length \(i\le t\) is at most \((2q(m+t+1))^i\). Summing over \(i=0,\ldots,t\), with \(i=0\) accounting for the empty script, gives the stated bound.
		
		Every word in \(\cB_t(x)\) can be obtained from \(x\) by at least one such script. Since distinct scripts may produce the same output word,
		\[
		|\cB_t(x)|
		\le \sum_{i=0}^{t}\bigl(2q(m+t+1)\bigr)^i
		\le (t+1)\bigl(2q(m+t+1)\bigr)^t
		=r_t(m)=\Theta_{q,t}(m^t).
		\]
	
	\section{Proof of Lemma~\ref{lem:short-edit}}\label{app:short-edit}
		Since \(r_{2t}(\ell)=\Theta_{q,t}(\ell^{2t})\),	exponential growth dominates \(r_{2t}(\ell)\), so the set defining \(\rho\) is nonempty for every \(h\ge1\).
		To bound its minimum, set
		\[
		\ell_h=h+\ceil{(2t+1)\log_q(h+1)}.
		\]
		Then
		\[
		q^{\ell_h-h}\ge(h+1)^{2t+1},
		\qquad
		r_{2t}(\ell_h)=O_{q,t}(h^{2t}).
		\]
		Thus, \(q^{\ell_h}\ge q^h r_{2t}(\ell_h)\) for all sufficiently large \(h\), giving
		\[
		h\le\rho\le\ell_h
		=h+O_{q,t}(\log(h+1)).
		\]
		The finitely many smaller values of \(h\) are covered by enlarging the implicit constant.
		
		Starting with \(\cA^\rho\), repeatedly select the lexicographically first remaining word and discard all remaining words within edit distance
		\(2t\) of it, until no words remain. The selected words have pairwise edit distance at least \(2t+1\) and hence form a \(t\)-edit-correcting code.
		
		By Lemma~\ref{lem:edit-degree}, each selection discards at most
		$
		r_{2t}(\rho)
		=(2t+1)\bigl(2q(\rho+2t+1)\bigr)^{2t}
		$
		words. The resulting code therefore has size at least
		\[
		\frac{q^\rho}{r_{2t}(\rho)}\ge q^h,
		\]
		where the last inequality follows from the definition of \(\rho\). Mapping the words in \(\cA^h\), in lexicographic order, to the first \(q^h\) selected codewords defines the required injective encoder \(E\).

	\section{Detailed complexity analysis}\label{app:runtime}
	
	We analyze deterministic implementations on a random-access	machine, charging arithmetic operations by their bit	complexity and memory accesses by the bit length of the address.	Throughout, \(q\) and \(t\) are fixed, and \(\widetilde O\)	suppresses polylogarithmic factors in \(n\). All public tables and working storage used below have size polynomial in \(n\), so their addresses have \(O_{q,t}(\log n)\) bits.
	
	Public setup is performed once and reused for subsequent operations. We first analyze the local graph routines and the short-code construction separately, and then combine their costs. The final encoding and decoding bounds include the one-time setup cost.
	
	\subsection{Graph coloring costs}\label{app:graph-costs}
	
	We begin with the construction in Section~\ref{sec:multi}.
	Let \(T_{\cN}\) denote the worst-case cost of enumerating the neighbors of a vertex, and let \(T_{\id}\) denote the cost of forming one \(K\)-bit identifier. Write \(T_{\mathrm{color}}\) for the cost of evaluating the coloring map \(f\) at one vertex, and \(T_{\mathrm{allocation}}\) for a common upper bound on the following two operations:
	selecting one assigned color pair \(\pi_v(j)\in S(v)\) by rank \(j\), and validating a proposed color pair while recovering its rank.
	
	We work in the parameter regime used by the edit-code construction in Section~\ref{sec:edit}, in which \(r\) and \(K\) are polynomially bounded in \(n\). Consequently,
	\[
	b:=\log_2 Q=O(\log n),\qquad
	B=O(\log n),\qquad
	\log L=O(\log n).
	\]
	
	\paragraph{Field and prime setup.}
	
	Represent \(\F_Q\) in a binary polynomial basis, using the coefficient vectors to identify its elements with \(\nz{Q}\). Shoup's deterministic algorithm \cite{shoup1990new} constructs an irreducible polynomial of degree \(b\) over \(\F_2\) in time polynomial in \(b\). Thus, constructing this representation of	\(\F_Q\) takes only polylogarithmic time in \(n\).
	
	The prime bound \eqref{eq:prime-bound} gives \(p_L=O(L\log L)\). Applying the sieve of Eratosthenes with a	doubled cutoff until at least \(L\) primes have been found produces the first \(L\) primes in \(\widetilde O(L)\) bit	operations. Hence, the combined field and prime setup cost is
	\begin{equation}\label{eq:public-setup}
		T_{\mathrm{setup}}=\widetilde O(L).
	\end{equation}
	
	\paragraph{Evaluating the coloring map.}
	
	To compute \(f(v)\), enumerate \(\cN(v)\), form the identifiers of \(v\) and its neighbors, and construct their identifier polynomials. Each polynomial has at most \(K\) coefficients.
	At a fixed field element, Horner's rule evaluates all these	polynomials using \(O((r+1)K)\) field operations.
	Scanning the \(Q\) field elements to find the first separating element therefore requires \(O(Q(r+1)K)\) field operations.
	
	Schoolbook arithmetic in the chosen polynomial basis uses \(O(b^2)\) bit operations per field operation. Since \(b=O(\log n)\), these arithmetic costs, together with the address costs, are absorbed by \(\widetilde O\). Thus,
	\begin{equation}\label{eq:color-time}
		T_{\mathrm{color}}
		\le T_{\cN}+(r+1)T_{\id}
		+\widetilde O\bigl(Q(r+1)K\bigr).
	\end{equation}
	
	\paragraph{Color pair selection and validation.}
	
	Both operations first enumerate \(\cN(v)\) and compute \(f(v)\) and \(f(u)\) for every \(u\in\cN(v)\).
	Each evaluation of \(f(u)\) uses the identifier polynomials	of \(u\) and its neighbors, but does not require their colors.
	Thus, these at most \(r+1\) color evaluations are \emph{not} recursive:	they access only identifiers of vertices at graph distance at most two from \(v\), without enumerating the full vertex set.
	
	Once these colors are available, scan the public list \(p_1,\ldots,p_L\) in increasing order, testing each prime for goodness with respect to \(v\).\footnote{The ordered set
		\(\good(v)\) is represented implicitly by these tests and need not be constructed or stored separately.}
	Selection stops when the \((j+1)\)-st good prime is encountered. To validate a proposed pair \((i,\eta)\), first check \(i\in\nz{L}\), then scan through index \(i\) to test goodness, check the residue, and recover the rank. The pair is accepted only if the recovered rank is less than \(J\).
	
	A full scan uses at most \(L(r+1)\) modular reductions and comparisons. Both the colors and the primes have \(O(\log n)\)-bit representations, so each such operation has polylogarithmic bit cost. Therefore,
	\begin{equation}\label{eq:allocation-time}
		T_{\mathrm{allocation}}
		\le T_{\cN}+(r+1)T_{\mathrm{color}}
		+\widetilde O\bigl(L(r+1)\bigr).
	\end{equation}
	This bound includes all neighborhood enumerations needed for the nested color evaluations.
	
	\paragraph{Specialization to the confusion graph.}
	
	For the edit-code construction, take \(r=r_{2t}(n)=\Theta_{q,t}(n^{2t})\).
	For any \(m\le n\), Lemma~\ref{lem:edit-degree} bounds the number of scripts of at most \(2t\) edits starting from a length-\(m\) input word by $r_{2t}(m)\le r$.
	To enumerate its neighbors, generate these scripts and retain their length-\(m\) outputs, excluding the input word itself.
	Each output can be materialized and filtered in	\(\widetilde O_{q,t}(n)\) time.
	Removing duplicates by sorting does not change the resulting bound:
	\begin{align}
		T_{\cN}=\widetilde O_{q,t}(rn).\label{eq:neighbors-cost}
	\end{align}
	
	The identifier
	$
	\id(x)=\sum_{\mu=1}^m x_\mu q^{m-\mu}
	$
	can be formed by Horner's rule in base \(q\).
	This uses at most \(n\) additions and multiplications by the fixed integer \(q\), on integers of \(O_q(n)\) bits. Hence,
	\begin{align}
		T_{\id}=\widetilde O_q(n^2).\label{eq:id-cost}
	\end{align}
	
	The parameter choices in Section~\ref{sec:edit} give
	\[
	K=O_q(n),\qquad
	Q=O_{q,t}(rn),\qquad
	B=O_{q,t}(\log n),\qquad
	L=O_{q,t}(r\log n).
	\]
	Substituting these estimates into \eqref{eq:neighbors-cost}, \eqref{eq:id-cost} and \eqref{eq:color-time} gives
	\[
	T_{\mathrm{color}}
	\le \widetilde O_{q,t}
	\bigl(rn+(r+1)n^2+r(r+1)n^2\bigr)
	=\widetilde O_{q,t}(r^2n^2).
	\]
	Equation~\eqref{eq:allocation-time} then yields
	\begin{equation}
		T_{\mathrm{allocation}}
		=\widetilde O_{q,t}(r^3n^2).\label{eq:edit-interface-costs}
	\end{equation}
	
	The integer parameters \(r,K,Q,B,L,J\), and then \(h_1,h_2\)	once \(p_L\) is available, can be computed in polylogarithmic	time in \(n\). Parameters involving floors or ceilings of logarithms can be computed exactly by generating successive integer powers of the relevant base and comparing them with the corresponding threshold. Since the bases are fixed and the thresholds are polynomially bounded in \(n\), these computations take polylogarithmic time in \(n\).
	It remains to compute \(\rho\), construct the short-code table, and account for its use.
	
	\subsection{Short-code costs}\label{app:short-costs}
	
	\paragraph{Determining the length.}
	
	Compute \(\rho\) by testing \(s=h,h+1,\ldots\) until
	\[
	q^{s-h}\ge r_{2t}(s)
	=(2t+1)\bigl(2q(s+2t+1)\bigr)^{2t}
	\]
	holds for the first time.
	Lemma~\ref{lem:short-edit} gives
	\[
	\rho=h+O_{q,t}(\log(h+1)),
	\]
	so this search uses only \(O_{q,t}(\log(h+1))\) trials.
	The integers used in these comparisons have	\(O_{q,t}(\log(h+1))\) bits. Thus, the length computation is negligible compared with the table construction below.
	
	Once \(\rho\) is known, the remaining lengths
	\[
	m=n-\rho,\qquad k=m+\log_q J
	\]
	are obtained by integer subtraction and addition.
	
	\paragraph{Constructing the encoder table.}
	
	Implement the greedy construction using a marking array indexed	by all \(q^\rho\) words in \(\cA^\rho\).
	Scan these words in lexicographic order. Whenever an unmarked word is encountered, select it and mark every length-\(\rho\) output of an edit script of length at most \(2t\) starting from
	it. Stop after \(q^h\) words have been selected. Map the messages in \(\cA^h\), in lexicographic order, to the selected words in order.
	
	The scan visits at most \(q^\rho\) words. By Lemma~\ref{lem:edit-degree}, the \(q^h\) selections generate at	most
	\[
	q^h r_{2t}(\rho)\le q^\rho
	\]
	edit scripts, where the inequality follows from the definition of \(\rho\).
	Each scan step and each generated script can be processed in \(O_{q,t}((\rho+1)^2)\) bit operations, including output construction, conversion to an array index, and memory access.
	The total construction cost is therefore
	\begin{align*}
		O_{q,t}(q^\rho(\rho+1)^2).
	\end{align*}
	To express this bound in terms of \(h\), note that
	\(r_{2t}(h)>1\), so \(\rho>h\). The minimality of \(\rho\)
	therefore gives
	\[
	q^\rho
	<q^{h+1}r_{2t}(\rho-1)
	=O_{q,t}\bigl(q^h(h+1)^{2t}\bigr),
	\]
	using \(\rho=O_{q,t}(h+1)\).
	Consequently, the entire table construction, including the length computation, takes
	\begin{equation}\label{eq:table-setup}
		T_{\mathrm{table}}
		=O_{q,t}\bigl(q^h(h+1)^{2t+2}\bigr)
	\end{equation}
	bit operations and
	\(q^h(h+1)^{O_{q,t}(1)}\) bits of space.
	The temporary marking array can be discarded after setup.
	
	\paragraph{Using the table.}
	
	To encode a word in \(\cA^h\), convert it to its base-\(q\)
	integer value and use that value as an index into the encoder
	table. Including conversion and output, this takes
	\begin{equation*}\label{eq:lookup-time}
		T_{\mathrm{lookup}}
		=O_{q,t}((h+1)^2).
	\end{equation*}
	
	For the decoder, we need exact membership testing and inversion of \(E\). The selected codewords are already sorted lexicographically. Binary search in this list uses \(O_q(h)\) comparisons of length-\(\rho\) words. If a match is found,	convert its table index back to a length-\(h\) word.
	Including address and conversion costs, this takes
	\begin{equation*}\label{eq:inversion-time}
		T_{\mathrm{inverse}}
		=O_{q,t}(h\rho)
		=O_{q,t}((h+1)^2).
	\end{equation*}
	Here, \(\rho\ge h\) absorbs the cost of converting the index.
	
	Serialization, inverse parsing, and conversion of \(j\in\nz{J}\) to or from its \(q\)-ary representation also take \(O_q((h+1)^2)\) bit operations by repeated multiplication or
	division by the fixed base \(q\).
	For the index conversion, we use \(\log_q J\le h\).
	
	In the edit-code construction, \(h,\rho=O_{q,t}(\log n)\). Thus, after table setup, all these
	short-code and representation operations take \(O_{q,t}((\log n)^2)\) time.
	
	\subsection{Overall encoding and decoding costs}
	\label{app:composition}
	
	We now combine the local graph routines with the short-code
	table.
	
	\paragraph{Public setup.}
	
	Let \(T_{\mathrm{public}}\) denote the complete one-time setup cost, including the integer parameters, the field representation, the prime list, and the short-code table.
	Equations~\eqref{eq:public-setup} and~\eqref{eq:table-setup} give
	\begin{equation*}\label{eq:total-public-setup}
		T_{\mathrm{public}}
		\le \widetilde O_{q,t}(L)
		+O_{q,t}\bigl(q^h(h+1)^{2t+2}\bigr).
	\end{equation*}
	The integer parameter computations are absorbed by these terms.
	
	For the edit-code parameters,
	\[
	L=\Theta_{q,t}(n^{2t}\log n),
	\qquad
	h=4t\log_q n+O_{q,t}(\log\log n).
	\]
	In particular, \(q^h=\widetilde O_{q,t}(n^{4t})\).
	It follows that
	\begin{align}
		T_{\mathrm{public}}
		=\widetilde O_{q,t}(n^{4t}).\label{eq:public-costs}
	\end{align}
	
	\paragraph{Encoding.}

	After setup, encoding requires selecting one color from \(S(x)\) by rank, serializing that color, applying the short-code encoder, and writing the length-\(n\) output.
	Charging the one-time setup to this operation gives
	\begin{equation*}\label{eq:encoder-total}
		T_{\mathrm{encoder}}
		\le T_{\mathrm{public}}
		+T_{\mathrm{allocation}}
		+O_{q,t}((h+1)^2)
		+\widetilde O_q(n).
	\end{equation*}
	Since \(r=\Theta_{q,t}(n^{2t})\),
	\eqref{eq:edit-interface-costs} gives
	\[
	T_{\mathrm{allocation}}
	=\widetilde O_{q,t}(n^{6t+2}).
	\]
	This term dominates the other costs, so
	\[
	T_{\mathrm{encoder}}
	=\widetilde O_{q,t}(n^{6t+2}).
	\]
	
	\paragraph{Candidate validation.}
	
	Let \(T_{\mathrm{validation}}\) denote the cost of testing one length-\(n\) candidate after setup.
	Accessing and splitting the candidate costs \(\widetilde O_q(n)\). Exact suffix inversion and parsing take \(O_{q,t}((h+1)^2)\) time.
	If these tests succeed, the allocation routine validates the proposed color and recovers its rank. Hence,
	\begin{equation}\label{eq:candidate-time}
		T_{\mathrm{validation}}
		\le \widetilde O_q(n)
		+O_{q,t}((h+1)^2)
		+T_{\mathrm{allocation}}
		=\widetilde O_{q,t}(n^{6t+2}).
	\end{equation}
	
	\paragraph{Decoding.}
	
	The initial length test rejects any received word whose length differs from \(n\) by more than \(t\).
	If the input length is not supplied, inspecting at most	\(n+t+1\) symbols suffices to perform this test, so its cost is \(\widetilde O_q(n)\).
	
	Now suppose that \(y\) passes the length test.
	Since \(|y|\le n+t\), the number of edit scripts of length at most \(t\) starting from \(y\) is
	at most
	\[
	r_t(n+t)=O_{q,t}(n^t).
	\]
	Each output has length \(O_{q,t}(n)\), so materializing all	outputs and retaining those of length \(n\) costs	\(\widetilde O_{q,t}(n^{t+1})\).
	
	Including public setup, the total decoding cost is therefore
	\begin{equation*}\label{eq:decoder-total}
		T_{\mathrm{decoder}}
		\le T_{\mathrm{public}}
		+\widetilde O_{q,t}(n^{t+1})
		+O_{q,t}(n^t)T_{\mathrm{validation}}.
	\end{equation*}
	Substituting the setup and validation bounds \eqref{eq:public-costs}, \eqref{eq:candidate-time} gives
	\[
	T_{\mathrm{decoder}}
	=\widetilde O_{q,t}(n^{7t+2}).
	\]
	
\end{document}